\pdfoutput=1

\documentclass[11pt]{article}
\usepackage[footnotesize]{caption}
\usepackage{geometry}
\usepackage{hyperref}
\usepackage{amsmath,amsfonts,amssymb,latexsym}
\usepackage{graphicx,subfig}
\usepackage{color}
\usepackage{cite}

\numberwithin{equation}{section} 
\numberwithin{figure}{section}
\numberwithin{table}{section} 
\begin{document}

\begin{titlepage}

\thispagestyle{empty}
\setcounter{page}{0}

\vskip 1.5cm

\begin{center}
{\LARGE\bf New perspectives on constant-roll inflation}

\vskip 2cm

{\large  Francesco Cicciarella${}^{a,}$\footnote{f.cicciarella1@gmail.com}, Joel Mabillard${}^{b,}$\footnote{joel.mabillard@ed.ac.uk}, Mauro Pieroni${}^{c,}$\footnote{mauro.pieroni@uam.es}}\\[3mm]
{\it{
${}^{a}$ Dipartimento di Fisica, Universit\`a di Pisa, Largo Bruno Pontecorvo 3, 56127 Pisa, Italy \\
${}^{b}$ School of Physics and Astronomy, University of Edinburgh, Peter Guthrie Tait Road, Edinburgh, EH9 3FD, United Kingdom \\
${}^{c}$ Instituto de F\'isica Te\'orica UAM/CSIC C/ Nicol\'as Cabrera 13-15 Universidad Aut\'onoma de Madrid Cantoblanco, Madrid 28049, Spain\\
}}
\end{center}

\vskip 1cm
\centerline{ {\bf Abstract}}{
We study constant-roll inflation using the $\beta$-function formalism. We show that the constant rate of the inflaton roll is translated into a first order differential equation for the $\beta$-function which can be solved easily. The solutions to this equation correspond to the usual constant-roll models. We then construct, by perturbing these exact solutions, more general classes of models that satisfy the constant-roll equation asymptotically. In the case of an asymptotic power law solution, these corrections naturally provide an end to the inflationary phase. Interestingly, while from a theoretical point of view (in particular in terms of the holographic interpretation) these models are intrinsically different from standard slow-roll inflation, they may have phenomenological predictions in good agreement with present cosmological data.}

\end{titlepage}
\tableofcontents

\section{Introduction}
\label{sec:introduction}
Nowadays inflation is widely accepted as a cornerstone of early time cosmology. However, while the main mechanism that drove this early phase of accelerated expansion starts to be understood, the definition of a concrete model that is completely satisfying from a theoretical point of view is still lacking. Since the original proposals~\cite{Guth:1980zm,Linde:1981mu,Albrecht:1982wi,Linde:1983gd,Starobinsky:1980te}, a huge amount of models to realize inflation has been proposed (for a fairly complete review see for example~\cite{Martin:2013tda}) and in some cases theoretical predictions are so close that they are nearly indistinguishable. A well known example of this degeneracy\footnote{However, it is fair to point out that some slight differences in the predictions for these models can be found by performing a more accurate analysis that keeps into account the physics of reheating~\cite{Bezrukov:2011gp}.} is the case of $R^2$ inflation~\cite{Starobinsky:1980te} and Higgs inflation~\cite{Bezrukov:2007ep,Bezrukov:2009db}.\\

In order to have a better connection between theory and experiments, several methods to classify inflationary models have been introduced over the last years~\cite{Mukhanov:2013tua,Roest:2013fha,Garcia-Bellido:2014gna}. In this context, we have proposed the $\beta$-function formalism for inflation~\cite{Binetruy:2014zya}. While at a formal level this characterization is equivalent to the standard procedure based on the specification of the inflationary potential, it typically permits to  perform exact computations even without assuming slow-roll. This formalism is inspired by a formal analogy between the equations describing the evolution of a scalar field in a Friedmann-Lema\^itre-Robertson-Walker (FLRW) background and a renormalization group equation (RGE). Because of this similarity, the evolution of the Universe during inflation is expressed in terms of a $\beta$-function similar to the one well known in the context of quantum field theory (QFT). When the magnitude of the cosmological $\beta$-function is smaller than one, the Universe experiences a phase of accelerated expansion and in particular it is easy to show that an exact de Sitter (dS) configuration is realized when the $\beta$-function is exactly equal to zero. As a consequence, in this framework inflation can be described using the Wilsonian picture of renormalization group (RG) flows between fixed points and it is thus natural to classify inflationary models in terms of a minimal set of parameters (\emph{i.e.} critical exponents) that specifies these flows. \\

The $\beta$-function formalism has a strong connection with the idea of applying holography to describe the inflationary Universe~\cite{Skenderis:2006jq,McFadden:2009fg,McFadden:2010na} which is nowadays a rapidly developing field of research (see for example~\cite{McFadden:2010vh,Bzowski:2012ih,Garriga:2014ema,Garriga:2014fda,Afshordi:2016dvb,Afshordi:2017ihr,Hawking:2017wrd,Conti:2017pqc}). In this framework, which is based on the (A)dS-CFT correspondence of Maldacena~\cite{Maldacena:1997re}, the deformation of an asymptotic (A)dS space-time (corresponding to the period inflation) is dual to the deformation of a (pseudo)-QFT (corresponding to the RG flow which is typically described in terms of a $\beta$-function). The cosmological $\beta$-function is thus interpreted as the usual $\beta$-function for the dual QFT. In fixed points with $\beta = 0 $ an exact dS configuration is realized and the dual QFT attains scale invariance becoming a CFT. A departure from these fixed points thus corresponds to scaling regions and on the cosmological side to inflationary epochs. This indicates that the appearance of a RG equation in the cosmological context is not fortuitous but rather supported by deeper theoretical motivations. \\

In this work, we study the case of constant-roll inflation in terms of the $\beta$-function formalism. Constant-roll inflation was originally introduced in~\cite{Martin:2012pe} and recently it attracted a lot of interest~\cite{Motohashi:2014ppa,Cai:2016ngx,Motohashi:2017aob,Odintsov:2017yud,Gao:2017uja,Gao:2017owg,Motohashi:2017vdc,Nojiri:2017qvx,Odintsov:2017qpp,Motohashi:2017gqb,Oikonomou:2017bjx} because of the possibility of predicting cosmological parameters that are in agreement with the most recent cosmological observational constraints~\cite{Ade:2015xua,Ade:2015lrj,Array:2015xqh}. In these models the scalar field $\phi$ is assumed to satisfy the constant-roll condition $\ddot{\phi} = - 3 \lambda H \dot{\phi}$, where $H$ is the Hubble parameter and $\lambda$ is some constant. Notice that for $\lambda \simeq 0 $ we recover the usual case of slow-roll inflation and for $\lambda = 1$ we have ultra slow-roll inflation~\cite{Tsamis:2003px,Kinney:2005vj,Namjoo:2012aa,Martin:2012pe,Dimopoulos:2017ged} (where the potential is exactly flat)\footnote{Interestingly, some recent works have shown that in this class of models the curvature perturbation on comoving slices and the curvature perturbation on uniform density slices do not coincide and are not conserved~\cite{Romano:2015vxz,Romano:2016gop}. Moreover, it has also been shown that it is possible to violate the non-Gaussianity consistency relations for single field inflationary models~\cite{Martin:2012pe,Namjoo:2012aa,Cai:2016ngx,Motohashi:2017aob,Odintsov:2017yud,Romano:2016gop,Mooij:2015yka}.}. We show that the class of models satisfying the constant-roll condition can be easily described in terms of the $\beta$-function formalism. Since the analysis carried out in this framework does not require the slow-roll approximation, it allows for a very simple characterization of constant-roll inflation\footnote{In fact some of the models that satisfy the constant-roll condition were already discussed in the general classification carried out in~\cite{Binetruy:2014zya}.}. Moreover, we show that with this approach we can define a further generalization of these models, leading to a new set of inflationary landscapes. In particular we construct a set of consistent inflationary models that asymptote to power-law inflation and have interesting cosmological predictions. \\

The paper is organized as follows. In Sec.~\ref{sec:general} we briefly review the $\beta$-function formalism and describe constant-roll inflation using this method. In Sec.~\ref{exact} we show that this approach is well suited to derive models for constant-roll inflation and discuss them in terms of the universality classes introduced in~\cite{Binetruy:2014zya}. In Sec.~\ref{sec:models} we go beyond the exact cases and construct quasi solutions of the constant-roll equation. We present their interpolating behavior and their phenomenological predictions. Our concluding remarks are given in Sec.~\ref{sec:conclusions}.

\section{Setting up the model}
\label{sec:general}
In this section we start by giving a brief review of the $\beta$-function formalism for models of inflation introduced in~\cite{Binetruy:2014zya} and then we discuss the case of the constant-roll within this framework. Although in this paper we consider a single scalar field minimally coupled to gravity and  with standard kinetic term, the results can be generalized to more complex scenarios (following the analysis of~\cite{Pieroni:2015cma,Binetruy:2016hna}).

\subsection{$\beta$-function formalism - A short review}
\label{subsec:General_beta_Function_Formalism}
The action for a homogeneous classical scalar field $\phi(t)$ minimally coupled to gravity in a FLRW universe with line element $\mathrm{d}s^2=-\mathrm{d}t^2+a^2(t)\mathrm{d}\vec{x}^2$ reads (in natural units $M_p^2\equiv (8\pi G)^{-1}=1$):
\begin{align}
	\label{eq:general_action}
	S&=\int \mathrm{d}^4 x \sqrt{-g}\left(\frac{ R }{2}-\frac{1}{2}\partial^\mu\phi\partial_\mu\phi-V(\phi)\right)\;.
\end{align}
Using the definition of the stress-energy tensor $T_{\mu\nu}\equiv -2(\delta S_m/\delta g^{\mu\nu})/\sqrt{-g}$, we can easily compute the pressure and the energy density of the scalar field:
\begin{align}
	p&=\frac{\dot{\phi}^2}{2}-V(\phi)\;,&& \rho=\frac{\dot{\phi}^2}{2}+V(\phi)\;.
\end{align}
The dynamical evolution of the system is then set by the Einstein equations which read:
\begin{align}
	\label{eq:general_EinsteinEquations}
	H^2&\equiv\left(\frac{\dot{a}}{a}\right)^2=\frac{\rho}{3}\;,&&	-2\dot{H}=p+\rho=\dot{\phi}^2\;.
\end{align}
Combining these equations we obtain the equation of state for the scalar field:
\begin{align}
	\label{eq:general_equationofstate}
	\frac{p+\rho}{\rho} = -\frac{2}{3}\frac{\dot{H}}{H^2}\;.
\end{align}
At this point, we can study the system in terms of the Hamilton-Jacobi approach of Salopek and Bond~\cite{Salopek:1990jq}. Assuming the dependence $\phi\equiv\phi(t)$ to be at least piecewise monotonic, we can invert it to get $t(\phi)$ and use the field as a clock to describe the evolution. By introducing the superpotential $W(\phi)\equiv-2H$, we obtain:
\begin{align}
	\dot{W}&=W_{,\phi}\dot{\phi}=-2\dot{H}=\dot{\phi}^2\;,&&\mbox{hence}&&\dot{\phi}=W_{,\phi}\;.
\end{align}
In analogy with the definition of the $\beta$-function in the context of QFT we define:
\begin{align}
	\label{eq:general_betafunction}
	\beta&=\frac{\mathrm{d}\phi}{\mathrm{d} \ln a}=\frac{\dot{\phi}}{H}=-2\frac{\dot{\phi}}{W}=-2\frac{W_{,\phi}}{W}\;,
\end{align}
which exhibits the standard form of a RG equation with $\phi$ playing the role of the renormalization constant and $a$ playing the role of the renormalization scale. To be precise, the analogy with the QFT $\beta$-function is not only at a formal level. Indeed it is possible to relate the cosmological $\beta$-function of Eq.~\eqref{eq:general_betafunction} with the $\beta$-function describing the RG flow induced by some scalar operator in the dual QFT. However, in order to properly set this correspondence, we have to specify a mapping between the bulk inflaton and the coupling in the dual QFT. In particular, this typically requires the specification of some renomalization condition which in principle may require a modification of the simple expression of $\beta$ in terms of $W$. While a detailed discussion of the holographic interpretation is beyond the scope of this work, an accurate discussion of this procedure can be found in~\cite{McFadden:2013ria}. \\

Eq.~\eqref{eq:general_equationofstate} implies that a phase of accelerated expansion is realized\footnote{More precisely, $\ddot{a}/a>0$ requires $|\beta(\phi)|<\sqrt{2}$. For simplicity, in the following we assume inflation to end at $|\beta(\phi)|\sim 1$.} for $| \beta (\phi) | \ll 1$. In analogy with the RG approach, we identify the zeros of the $\beta$-function as fixed points, which in the cosmological case correspond to exact dS solutions. Depending on the sign of the $\beta$-function inflationary periods are then represented by the flow of field away (or towards) these fixed points\footnote{\label{flow_towards}When the flow is towards the fixed point there is no natural end to the period of accelerated expansion. Clearly, this configuration is not suitable to describe inflation. However, as discussed in~\cite{Cicciarella:2016dnv}, this scenario can be relevant in the description of the late time acceleration of the Universe in terms of quintessence.}. As a consequence, it is possible to classify the various models of inflation according to the behavior of the $\beta$-function in the neighborhood of the fixed point, rather than according to the potential. The advantage of this approach is that specifying a form for $\beta(\phi)$ actually defines an universality class that encompasses more models which might have in principle very different potentials but nevertheless yield to a similar cosmological evolution (and thus to similar predictions for cosmological observables such as the scalar spectral index $n_s$ and the tensor-to-scalar ratio $r$). Notice that in order to realize a phase of accelerated expansion we only need $| \beta (\phi) | \ll 1$ and in general this does not require $\beta \rightarrow 0$. In particular, inflation can be realized even if $\beta(\phi)$ approaches a small constant value. As discussed in~\cite{Binetruy:2014zya}, this is the case of power law inflation~\cite{Lucchin:1984yf}. \\

In the $\beta$-function formalism the number of e-foldings $N$ can be expressed as:
\begin{align}
	\label{eq:general_numberofoefoldings}
	N=-\ln(a/a_{\mathrm{f}})=-\int_{\phi_\textrm{f}}^\phi\frac{\mathrm{d}\phi'}{\beta(\phi')}\;,
\end{align}
where $\phi_{\mathrm{f}}$ is the value of the field at the end of inflation. Similarly we can compute the superpotential:
\begin{equation}
	\label{eq:general_superpot}
	W(\phi) = W_{\textrm{f}} \exp \left( -\int_{\phi_\textrm{f}}^\phi \frac{\beta(\phi')}{2} \mathrm{d}\phi' \right)\;,
\end{equation}
and the inflationary potential:
\begin{align}
	\label{eq:general_potential}
	V(\phi)&=\frac{3}{4}W^2-\frac{1}{2}W_{,\phi}^2=\frac{3}{4}W^2\left(1-\frac{\beta^2(\phi)}{6}\right)\;,
\end{align}
whose parameterization is similar to the one in the context of supersymmetric quantum mechanics (for reference see for example~\cite{Binetruy:2006ad}). It is important to stress that so far all the computations are exact, \emph{i.e.} we have not performed any approximation and the analysis holds even if we are not assuming slow-roll.\\

Assuming now to be in a neighborhood of the fixed point, we have $|\beta(\phi)|\ll 1$ and $n_s$ and $r$ at the lowest order in terms of $\beta$ and its derivative simply read:
\begin{align}
	\label{eq:general_nsandr}
	n_s-1&\simeq-\left(\beta^2+2\beta_{,\phi}\right)\;,&&r=8\beta^2\;.
\end{align}
In order to obtain the standard expressions of $n_s$ and $r$ in terms of $N$, we should first determine the value of $\phi$ at the end of inflation (using the condition $|\beta(\phi_{\mathrm{f}})|\sim 1$). We should then proceed by computing $N(\phi)$ (using Eq.~\eqref{eq:general_numberofoefoldings}) and invert it into $\phi(N)$ to express $n_s$ and $r$ in terms of the number of e-foldings.

\subsection{Constant-roll inflation}
\label{subsec:General_Constant_Roll_Inflation}
Models of constant-roll inflation have been studied in the last years as an alternative that goes beyond the simple slow-roll realization of inflation. Unlike the usual slow-roll approach, the second derivative $\ddot{\phi}$ in the Klein-Gordon equation is not neglected in constant-roll inflation. A constant rate of rolling for the inflaton field, \emph{i.e.} $\ddot{\phi}/(H\dot{\phi})$ is then assumed. The first case to be studied was the so called \emph{ultra slow-roll inflation}~\cite{Martin:2012pe}, in which $\ddot{\phi}/(H\dot{\phi})=-3$, corresponding to an exactly flat potential $\partial V/\partial\phi=0$. Later, deviation from this regime have been considered in~\cite{Motohashi:2014ppa,Cai:2016ngx,Motohashi:2017aob,Odintsov:2017yud,Gao:2017uja,Gao:2017owg,Motohashi:2017vdc,Nojiri:2017qvx,Odintsov:2017qpp,Motohashi:2017gqb,Oikonomou:2017bjx}, with $\ddot{\phi}/(H\dot{\phi})=\mbox{const}\ne -3$.\\

In general the constant-roll condition\footnote{Note that $\lambda$ here is related to $\alpha$ of~\cite{Motohashi:2014ppa} as $\lambda=1+\alpha/3$ and to $\beta$ of~\cite{Motohashi:2017aob} as $\lambda=-\beta/3$.} is:
\begin{equation}
	\label{eq:constant_roll_definition}
	\ddot{\phi} = -3 \lambda H \dot{\phi}\;.
\end{equation}
In order to express it in terms of the $\beta$-function formalism we should start by expressing $\ddot{\phi}$ in terms of the superpotential ($\dot{\phi}=W_{,\phi}$):
\begin{align}
	\ddot{\phi}&=\dot{W_{,\phi}}=W_{,\phi\phi}\dot{\phi}\;,
\end{align}
so that the constant-roll equation~\eqref{eq:constant_roll_definition} becomes:
\begin{equation}
	\label{eq:w_phiphi}
	W_{,\phi\phi}=\frac{3\lambda}{2}W \;.
\end{equation}
At this point we compute the first derivative of the $\beta$-function:
\begin{align}
	\beta_{,\phi}&=-\frac{2W_{,\phi\phi}}{W}+2\left(\frac{W_{,\phi}}{W}\right)^2\;,&&\mbox{yielding}&&W_{,\phi\phi}=\frac{W}{2}\left(\frac{1}{2}\beta^2-\beta_{,\phi}\right)\,,
\end{align}
to express Eq.~\eqref{eq:w_phiphi} as:
\begin{align}
	\label{eq:constant_roll_beta}
	\frac{1}{2}\beta^2-\beta_{,\phi}&=3\lambda\;,
\end{align}
which is a nonlinear Riccati equation (the corresponding linear second-order ODE is Eq.~\eqref{eq:w_phiphi}). This equation shows a first advantage of this formalism: instead of dealing with a second order differential equation for $\phi$ in which in addition we need to specify $H(t)$, here we have a first order differential equation for $\beta(\phi)$ that can be easily integrated once the value of the constant $\lambda$ is chosen. \\

The scalar spectral index and the tensor-to-scalar ratio from Eq.~\eqref{eq:general_nsandr} are given in this particular case by:
\begin{align}
	\label{eq:ns_r_constant_roll}
	n_s-1&=-\left(\beta^2+2\beta_{,\phi}\right)=-2\beta^2+6\lambda=6\lambda-\frac{r}{4}\;.
\end{align}
We see that $n_s$ and $r$ are not independent. This should not come as a surprise, as at the lowest order the two parameters depends uniquely on $\beta$ and $\beta_{,\phi}$ and the constant-roll condition actually establishes a relationship between them.

\section{Exact solutions}
\label{exact}
In this section we study the exact solutions of Eq.~\eqref{eq:constant_roll_beta}. Depending on the sign of the constant $\lambda$, we find different parameterizations for the $\beta$-functions. In particular we show that we can easily recover all the known solutions of~\cite{Motohashi:2014ppa}. We discuss these solutions in terms of the universality classes introduced in~\cite{Binetruy:2014zya} and explain that some of these cases should be interpreted as composite models that interpolate between two different classes. Interestingly some of the cases discussed in this section have been already considered in~\cite{Binetruy:2014zya} illustrating the generality of the results obtained using this formalism (which indeed does not require slow-roll). 

\subsection{Exact dS and chaotic models - $\lambda=0$ }
\label{subsubsec:General_Exact_solution_exact_chaotic}
We start by considering the special case\footnote{This case does not properly belong to the constant-roll class of models, as from Eq.~\eqref{eq:constant_roll_definition} it follows $\ddot{\phi}=0$, which is the standard slow-roll condition.} $\lambda=0$ so that the equation for the $\beta$-function becomes $\beta_{,\phi}=\beta^2/2$. It is easy to show that this differential equation has two solutions:
\begin{equation}
 	\beta(\phi)=0 \; , \qquad \qquad \beta(\phi)= -\frac{2}{\phi} \; .
 \end{equation} 
The first case corresponds to an exact de Sitter configuration in which $a\propto e^{Ht}$, with constant $H$. This case is not dynamical (there is no flow and the condition $\beta\sim 1$ is never attained \emph{i.e.} there is no end to inflation) and it cannot be used to describe the inflationary universe. In particular, some mechanism must be introduced in order to end the period of accelerated expansion. In terms of the $\beta$-function formalism this can be implemented by introducing one (or more terms) in the parameterization of $\beta$ which induces the flow. On the other hand, the second case is the $\beta$-function for the chaotic class \textbf{Ib(1)} (of~\cite{Binetruy:2014zya}) with $\beta_1=2$ that corresponds to the case of Chaotic inflation~\cite{Linde:1983gd}. In general this class comprehends inflationary models for which the fixed point is reached at $|\phi|\to\infty$ and includes, for generic $\beta_1$, all the inflationary potentials of the form $V(\phi)=V_{\textrm{f}} \, \phi^{\beta_1}$. \\

Let us consider the generic case with $\beta_1$. The number of e-foldings is easily computed:
\begin{equation}
	N=\frac{\phi^2-\phi_{\mathrm{f}}^2}{2\beta_1}\;,
\end{equation}
with $|\phi_{\mathrm{f}}|=\beta_1$. We obtain $\phi(N)=\sqrt{2\beta_1 N+\beta_1^2}$ and thus:
\begin{equation}
	\beta(N)=-\frac{\beta_1}{\sqrt{2\beta_1N+\beta_1^2}}\;.
\end{equation}
Setting $\beta_1=2$, we have for $n_s$ and $r$:
\begin{align}
	\label{eq:ns_and_r_chaotic}
	n_s-1\simeq-\frac{2}{N}\;,&& r\simeq\frac{8}{N}\;,
\end{align}
which are exactly the predictions given by the standard chaotic model with quadratic potential~\cite{Linde:1983gd}.

\subsection{Constant $\beta$-function - $\lambda > 0$}
\label{subsubsec:General_Exact_solution_cst}
We now consider the case of a constant solution for Eq.~\eqref{eq:constant_roll_beta}, i.e. $\beta_{,\phi}=0$. Then $\beta(\phi)=\pm\sqrt{6\lambda}$, implying $\lambda>0$. As pointed out in~\cite{Binetruy:2014zya}, this case simply corresponds the power law class \textbf{Ib(0)}. Notice that Power law inflation~\cite{Lucchin:1984yf} is a static solution (different from dS) of Einstein Equations and thus there is no natural end to inflation. Similarly to the case of an exact dS configuration, in order to define a complete model it is necessary to introduce a mechanism to leave the period of accelerated expansion. As later shown in Sec.~\ref{sec:models}, such a mechanism might be provided by the insertion of some corrections to the exact solution outlined in this section.\\

In the case of power law solutions the scale factor is not exponentially increasing with time but rather scales with some power of the cosmic time. In the present case it is easy to show that $a\sim t^{1/(3\lambda)}$. As the $\beta$-function does not depend on $\phi$, the same will be true for $n_s$ and $r$, which can then immediately be computed:
 \begin{align}
	n_s-1=-\beta^2=-6\lambda=-\frac{r}{8}\;,&&r=48\lambda\;.
\end{align}
These models are strongly disfavored by the latest cosmological data~\cite{Ade:2015xua,Ade:2015lrj,Array:2015xqh}.

\subsection{Hyperbolic tangent and cotangent - $\lambda>0$ }
\label{subsubsec:General_Exact_solution_tanh}
We now solve~\eqref{eq:constant_roll_beta} for positive $\lambda$ and $\beta_{,\phi}\ne 0$ to find two possible solutions:
\begin{align}
	\label{eq:constant_roll_tanhbeta_and_cotanhbeta}
	\beta(\phi)&=\sqrt{6\lambda}\tanh\left(-\sqrt{\frac{3\lambda}{2}} \phi \right)\;,&&\text{and}&&\beta(\phi)=\sqrt{6\lambda}\ \text{cotanh}\left(-\sqrt{\frac{3\lambda}{2}} \phi \right)\;.
\end{align}
Let us start by considering the first of these two solutions. In this case the functional form for $\beta(\phi)$ clearly presents a zero in $\phi = 0$. We proceed with our analysis by computing the number of e-foldings:
\begin{align}
	\label{eq:N_expression_hyperbolic}
	N&=-\int_{\phi_\textrm{f}}^\phi\frac{\mathrm{d}\phi'}{\beta(\phi')}=\frac{1}{6\lambda}\ln\left[\frac{\sinh^2\left(-\sqrt{\frac{3\lambda}{2}} \phi \right)}{\sinh^2\left(-\sqrt{\frac{3\lambda}{2}} \phi_\textrm{f} \right)}\right]\;.	
\end{align}
To have a positive number of e-folding (corresponding to an expanding Universe) we need $|\phi_f|<|\phi|$. This corresponds to a flow towards the fixed point. As already mentioned (see footnote~\ref{flow_towards}), this is not an appropriate description of the early Universe since there is no natural end to the period of inflation. Conversely, in this case a flow away from the fixed point would describe a shrinking Universe that of course is not relevant for the analysis carried out in this work.\\

For the second solution the fixed point is attained in the limit $\phi\rightarrow-\infty$. As a first step we obtain $\phi_\textrm{f}$ from $\beta(\phi_\textrm{f})=1$:
\begin{align}
	\label{eq:constant_roll_cosh2phif}
	\sqrt{6\lambda}\text{cotanh}\left(-\sqrt{\frac{3\lambda}{2}} \phi_\textrm{f} \right)&=1\;,&&\mbox{hence}&&
	\cosh^2\left(-\sqrt{\frac{3\lambda}{2}} \phi_\textrm{f} \right)=\frac{1}{1-6\lambda}\;. 
\end{align}
Notice that as $\lambda$ is positive in order for the field to be real valued at the end of inflation we need $0<\lambda<1/6$. As usual, we proceed by deriving the number of e-foldings:
\begin{align}
	\label{eq:constant_roll_Ncotanh}
	N&=\frac{1}{6\lambda}\ln\left[\frac{\cosh^2\left(-\sqrt{\frac{3\lambda}{2}} \phi \right)}{\cosh^2\left(-\sqrt{\frac{3\lambda}{2}} \phi_\textrm{f} \right)}\right] =\frac{1}{6\lambda}\ln\left[(1-6\lambda)\cosh^2\left(-\sqrt{\frac{3\lambda}{2}} \phi \right)\right] \;,	
\end{align}
where we used Eq.~\eqref{eq:constant_roll_cosh2phif}. Notice that since $\cosh^2\left(-\sqrt{\frac{3\lambda}{2}} \phi \right)>\cosh^2\left(-\sqrt{\frac{3\lambda}{2}} \phi_\textrm{f} \right)$ for $|\phi|>|\phi_f|$, $N$ is positive. At this point we can express $\beta$ in terms of $N$ as:
\begin{align}
	\beta(N)&=\sqrt{\frac{6\lambda}{1-(1-6\lambda)e^{-6\lambda N}}}\;.
\end{align}
We then find for the cosmological observables $n_s$ and $r$:
\begin{align}
	n_s-1=6\lambda\left(1-\frac{2}{1-(1-6\lambda)e^{-6\lambda N}}\right)\;, \qquad r=8\beta^2=\frac{48\lambda}{1-(1-6\lambda)e^{-6\lambda N}}\;.
\end{align}
For completeness, we report the scalar potential associated with this parameterization of the $\beta$-function:
\begin{align}
	V(\phi)&=V_f\left[-\lambda+(1-\lambda)\sinh^2\left(-\sqrt{\frac{3\lambda}{2}} \phi \right)\right]\;,
\end{align}
where $V_\textrm{f}$ is the normalization of the inflationary potential that as usual can be set using the COBE normalization~\cite{Ade:2015xua,Ade:2015lrj}.\\

We can appreciate the interpolating behavior of $\beta(N)$ (between the power law \textbf{Ib(0)} and the chaotic class \textbf{Ib(1)})  by taking first the limit $6\lambda N\gg1$:
\begin{align}
	\beta(N)&=\sqrt{\frac{6\lambda}{1-(1-6\lambda)e^{-6\lambda N}}}\simeq\sqrt{6\lambda}\;,
\end{align}
which is the power law class \textbf{Ib(0)} with $\beta_0=-\sqrt{6\lambda}$. On the other hand for $6\lambda N\ll 1$, we have:
\begin{align}
	\beta(N)&=\sqrt{\frac{6\lambda}{1-(1-6\lambda)e^{-6\lambda N}}}\simeq\sqrt{\frac{6\lambda}{1-(1-6\lambda)(1-6\lambda N)}} \simeq \frac{1}{\sqrt{N+1}}\;,
\end{align}
corresponding to the chaotic class \textbf{Ib(1)} with $\beta_1=2$, for which $n_s$ and $r$ are those given in Eq.~\eqref{eq:ns_and_r_chaotic}.

\subsection{Tangent - $\lambda<0$ }
\label{subsubsec:General_Exact_solution_tan}
Looking for a solution of Eq.~\eqref{eq:constant_roll_beta} with negative $\lambda$ and $\beta_{,\phi}\ne 0$, we find\footnote{Note that also $\beta(\phi)=\sqrt{6|\lambda|}\cot\left(-\sqrt{\frac{3|\lambda|}{2}}\phi\right)$ is a solution in this case. However, this reduces to Eq.\eqref{eq:constant_roll_tanbeta} through the redefinition $\phi\to\frac{\pi}{\sqrt{6|\lambda|}}-\phi $.}:
\begin{align}
	\label{eq:constant_roll_tanbeta}
	\beta(\phi)&=\sqrt{6|\lambda|}\tan\left(\sqrt{\frac{3|\lambda|}{2}} \phi \right)\;.
\end{align}
This form of $\beta$-function corresponds to the interpolating class introduced in~\cite{Binetruy:2014zya} with the choice $f=1/\sqrt{6|\lambda|}$. It interpolates between the linear \textbf{Ia(1)} and the chaotic class \textbf{Ib(1)}, respectively in the limits $\phi\rightarrow0$ and $\phi\rightarrow\frac{\pi}{\sqrt{6|\lambda|}}$.\\

The number of e-foldings associated with Eq.~\eqref{eq:constant_roll_tanbeta} is given by:
\begin{align}
	\label{eq:N_expression_tangent}
	N&=-\frac{1}{6|\lambda|}\ln\left[\frac{\sin^2\left(\sqrt{\frac{3|\lambda|}{2}} \phi \right)}{\sin^2\left(\sqrt{\frac{3|\lambda|}{2}} \phi_\textrm{f} \right)}\right]\;.	
\end{align}
We obtain $\phi_\textrm{f}$ from $\beta(\phi_\textrm{f})=1$ (in this case the field is positive, $\phi_\textrm{f}>\phi\geq0$):
\begin{align}
	\label{eq:constant_roll_sinh2phif}
	\sqrt{6|\lambda|}\tan\left(\sqrt{\frac{3|\lambda|}{2}} \phi_\textrm{f} \right)&=1\;,&&\mbox{hence}&&
	\sin^2\left(\sqrt{\frac{3|\lambda|}{2}} \phi_\textrm{f} \right)=\frac{1}{1+6|\lambda|}\;. 
\end{align}
Then substituting into Eq.~\eqref{eq:N_expression_tangent}:
\begin{align}
	\label{eq:constant_roll_Ntan}
	N&=-\frac{1}{6|\lambda|}\ln\left[(1+6|\lambda|)\sin^2\left(\sqrt{\frac{3|\lambda|}{2}} \phi \right)\right]\;,
\end{align}
and thus we get:
\begin{align}
	\beta(N)&=\sqrt{\frac{6|\lambda|}{(1+6|\lambda|)e^{6|\lambda| N}-1}}\;,
\end{align}
where we used that $\sin^2\left(\sqrt{\frac{3|\lambda|}{2}} \phi \right)=e^{-6|\lambda| N}/(1+6|\lambda|)$. We then find for $n_s$ and $r$:
\begin{align}
	n_s-1& \simeq -6|\lambda|-\frac{12|\lambda|}{(1+6|\lambda|)e^{6|\lambda| N}-1}\;, \qquad r =\frac{48|\lambda|}{(1+6|\lambda|)e^{6|\lambda| N}-1}\;.
\end{align}
The scalar potential associated with this $\beta$-function reads:
\begin{align}
	V(\phi)&=V_f\left[-|\lambda|+(1+|\lambda|)\cos^2\left(\sqrt{\frac{3|\lambda|}{2}} \phi \right)\right]\;,
\end{align}
where again $V_\textrm{f}$ is the normalization of the inflationary potential.\\

We can appreciate the interpolating behavior of this form of $\beta(N)$ by taking first the limit $6|\lambda| \gg1$ :
\begin{align}
	\beta(N)&= e^{-3|\lambda| N}\;,
\end{align}
which is the linear class \textbf{Ia(q)} with $\beta_1=3|\lambda|$, $q=1$. In this limit (\emph{i.e.} the small field limit of~\eqref{eq:constant_roll_tanbeta}) we get:
\begin{align}
n_s-1&\simeq -6|\lambda|\;,&& r\simeq 8e^{-6|\lambda|N}\;.
\end{align}
On the other hand for $6|\lambda| N\ll 1$, we have:
\begin{align}
	\beta(N)&\simeq\sqrt{\frac{6|\lambda|}{(1+6|\lambda|)(1+6|\lambda|N)-1}}=\sqrt{\frac{1}{N+1+6|\lambda| N}}\simeq\sqrt{\frac{1}{N+1}}\;,
\end{align}
corresponding to the chaotic class \textbf{Ib(1)} with $\beta_1=2$, for which $n_s$ and $r$ are those given in Eq.~\eqref{eq:ns_and_r_chaotic}. Recall that this is the large field limit, with $\sqrt{6|\lambda|} \phi \rightarrow \pi$.

\section{Beyond exact solutions}
\label{sec:models}
Having established that the $\beta$-function formalism is convenient to describe constant-roll inflation, in this section we show that this approach is also well suited for going beyond the exact cases discussed so far. We consider quasi solutions, \emph{i.e.} models that satisfy the constant-roll equation asymptotically (deep in the inflationary phase) but not at later times (towards the end of inflation). These approximated solutions are interesting for several reasons:
\begin{itemize}
	\item From a purely phenomenological point of view it is interesting to explore the possibility of defining a new set of models that may lead to interesting predictions for $n_s$ and $r$.
 	\item From a theoretical point of view, these models are (as we will explain in the following sections) intrinsically different from the usual models.  
 	\item As already mentioned in Sec.~\ref{subsubsec:General_Exact_solution_cst}, in the case of the power-law class \textbf{Ib(0)} (\emph{i.e.} power-law inflation) the model is incomplete as it lacks a method to put an end to the inflationary stage. The introduction of corrections is thus necessary to define a consistent model for inflation (\emph{i.e.} a graceful exit).
 \end{itemize} 
 Before considering some explicit models, let us first discuss the general procedure to implement this mechanism.\\

Let $F(\phi)$ be an exact solution of Eq.~\eqref{eq:constant_roll_beta} (\emph{i.e.} one of the functions discussed in the previous section) and let us express $\beta$-function as:
\begin{equation}
	\beta(\phi) = F(\phi) + f(\phi) \;,
\end{equation}
where $f(\phi)$ is a generic function of $\phi$. In order for $\beta(\phi)$ to asymptotically satisfy the constant-roll condition~\eqref{eq:constant_roll_beta}, the function $f(\phi)$ must satisfy:
\begin{equation}
	-f_{,\phi}(\phi) + f^2(\phi)/2 + f(\phi) F(\phi) \simeq 0 \; ,  
\end{equation}
as we go deeper into the inflationary stage. In particular this means that in this regime both  $f(\phi)$ and its first derivative go to zero. Notice that since $\beta$ is not an exact solution of~\eqref{eq:constant_roll_beta}, the prediction for $n_s$ and $r$ are not given by the special case~\eqref{eq:ns_r_constant_roll} but by the general expression~\eqref{eq:general_nsandr}.\\

For example, assuming $F(\phi)$ to have the form discussed in Sec.~\ref{subsubsec:General_Exact_solution_tan} (\emph{i.e.} the tangent case with $\lambda < 0$), we can express the $\beta$-function as: 
\begin{equation}
	\label{eq:deformed_tanget}
	\beta(\phi) = \sqrt{6|\lambda|}\tan\left(\sqrt{\frac{3|\lambda|}{2}} \phi \right) + f(\phi)  \; .
\end{equation}
At this point we can use the equations introduced in Sec.~\ref{subsec:General_beta_Function_Formalism} to study the deformation induced by the function $f(\phi)$ around the exact solution discussed Sec.~\ref{subsubsec:General_Exact_solution_tan}. While in the case defined by Eq.~\eqref{eq:deformed_tanget} the equations are not easy to be handled, an analytical treatment can be carried out for $F(\phi) = \pm \sqrt{6 \lambda}$ allowing for a systematical study of the deformations over the asymptotical power law solution. In the remaining of this section, we will focus on this case and we will explicitly construct some models that arise as deformation of the constant solution studied in Sec.~\ref{subsubsec:General_Exact_solution_cst} and discuss their predictions.

\subsection{Asymptotical power law solutions}
\label{sec:positive}
As already mentioned in Sec.~\ref{subsubsec:General_Exact_solution_cst}, an exactly constant $\beta$-function corresponds to (eternal) power law inflation and therefore is not a complete model. Adding contributions to the $\beta$-function leads to the definition of a consistent effective model that may naturally provide an end for inflation. Let us express the $\beta$-function as\footnote{To simplify the discussion we have chosen to explain the general picture in terms of a positive $\beta$-function, however, similar arguments apply for a negative $\beta(\phi)$.}:
\begin{equation}
	\label{eq:beta_asympt_pl}
	\beta(\phi) = \sqrt{6\lambda} + f(\phi) \; .
\end{equation}
Since $f(\phi)$ vanishes as we go deeper in the inflationary stage, in this regime it is the constant term that dominates giving the asymptotical power law solution. However, as we depart from this configuration, the function $f(\phi)$ (that here is chosen to be positive), eventually becomes sufficiently large (at $\beta(\phi_{\mathrm{f}}) \simeq 1$) and puts an end to inflation. \\

Before discussing the predictions associated with these classes of models and defining some examples where the situation depicted in the previous paragraph is realized, we need to stress some points:
\begin{itemize}
	\item These classes of models are intrinsically different from the classes introduced in~\cite{Binetruy:2014zya}. The presence of a constant term in the $\beta$-function implies that going deeper in the inflationary stage the $\beta$-function does not approach the usual dS configuration but rather a power law solution as the one discussed in Sec.~\ref{subsubsec:General_Exact_solution_cst}. 
	\item As $\beta(\phi)$ does not approach zero during the inflation, Eq.~\eqref{eq:general_numberofoefoldings} directly implies that in general\footnote{It is fair to point out that this can happen even if $\beta$ goes to zero. However, in the well-defined cases considered in~\cite{Binetruy:2014zya} $N$ is always diverging approaching the fixed point.} there could be an upper bound for $N$. As a consequence, in this case it is necessary to check that the deformation allows for a sufficiently long period of inflation. 
	\item As during inflation the $\beta$-function asymptotes to a constant, the dual theory does not attain conformal invariance. As a matter of fact, the constant term corresponds to a relevant operator which modifies the IR behavior (on the other hand the function $f(\phi)$ corresponds to the introduction of an irrelevant operator which modifies the UV behavior) of the RG flow. The QFTs that are dual to this class of models are not CFTs but rather theories with generalized conformal invariance~\cite{Jevicki:1998ub}. These are a subclass of hyperscaling violating theories~\cite{Charmousis:2010zz,Gouteraux:2011ce,Huijse:2011ef} where the dynamical critical exponent $z=1$ but $\theta \neq 0$ ($z \neq 1 $ leads to anistropic solutions which are beyond the interest of this paper). For a discussion of the holographic interpretation of this class of theories see for example~\cite{Jevicki:1998ub,Kanitscheider:2008kd,McFadden:2010na,Dong:2012se,Alishahiha:2012cm,Gath:2012pg,Hartnoll:2016apf}.
\end{itemize}

In order to compute the predictions for $n_s$ and $r$ associated with these classes of models we can directly substitute the parameterization of Eq.~\eqref{eq:beta_asympt_pl} into Eq.~\eqref{eq:general_nsandr} to get:
\begin{align}
	n_s-1&=-\left[6\lambda+2\sqrt{6\lambda}f+f^2+2f_{,\phi}\right]\;,\\
	r&=8\left[6\lambda+2\sqrt{6\lambda}f+f^2\right]\;.
\end{align}
We can directly see from these equations that the predictions for a $\beta$-function of form~\eqref{eq:beta_asympt_pl}, interpolate between the predictions of the power law class \textbf{Ib(0)} and of the class in which $f(\phi)$ belongs to. This mechanism of interpolation is analogous to the one discussed in~\cite{Binetruy:2014zya,Pieroni:2015cma} and in this case is clearly mediated by the value of $\lambda$. Indeed if $f,f_{,\phi}\ll\sqrt{6\lambda}$ at $N=60$, we recover the predictions of a power law model. On the other hand, if $f,f_{,\phi} \gg\sqrt{6\lambda}$ at $N=60$, the constant does not play a significant role and the predictions are the same as if the $\beta$-function were simply $\beta(\phi)=f(\phi)$. Notice however, that independently on the predictions at $N\simeq 60$, the presence of the (small) constant is still deeply affecting the asymptotic behavior according to the discussion of the previous paragraph.\\

Finally, let us define a minimal set of relevant examples. In order to produce a systematic analysis of the different possibilities, we can consider separately the following cases:
\begin{itemize}
	\item $f_{,\phi}/f \simeq 1$, functions of the Exponential class \textbf{II($\gamma$)}:\\
	$f(\phi)=e^{\gamma\phi}$ with $\gamma > 0$ and $\phi < 0$. \\
	The asymptotic power law solution is recovered for $\phi \rightarrow -\infty$.
	\item  $f_{,\phi}/f \ll 1$, functions\footnote{In principle we could also use functions of the Fractional class \textbf{Ib(p)} (with $p<1$) but as they are not leading to interesting predictions for $n_s$ and $r$ we are not considering them in our analysis.} of the Inverse \textbf{Ib(p)} and Chaotic class \textbf{Ib(1)}:\\
	$f(\phi)=\alpha\phi^{-n}$ with $\phi<0 $ and either $\alpha>0$ and $n$ even or $\alpha<0$ and $n$ odd.\\
	The asymptotic power law solution is recovered for $\phi \rightarrow -\infty$.
	\item   $f_{,\phi}/f \gg 1$, functions of the Monomial class \textbf{Ia(q)}:\\
	 $f(\phi)=\alpha\phi^n$ with  $\alpha>0$ and $\phi>0$.\\
	The asymptotic power law solution is recovered for $\phi \rightarrow 0$.
\end{itemize}
To be consistent with the convention used in~\cite{Binetruy:2014zya,Pieroni:2015cma}, in the following we proceed by choosing the sign of the terms appearing in the $\beta$-function in order to have always $\phi>0$ and positive constants. Clearly the results are independent on this choice.

\subsubsection{Exponential}
\label{sec:exponential}
Let us consider a $\beta$-function of the form:
\begin{equation}
	\label{eq:exponential_class}
	\beta(\phi)=-\sqrt{6\lambda}-\alpha e^{-\gamma\phi}\;,
\end{equation}
with $\alpha > 0$, $\gamma> 0$ and $\phi >0$ so that inflation is realized for large positive values of $\phi$. In this case the $\beta$-function goes from $-\sqrt{6\lambda}$ to $-1$ as $\phi$ decreases from $\phi > \phi_{\mathrm{f}}$ to $\phi_{\mathrm{f}}$. Notice that by redefining the inflaton field as $\phi\to\phi-\ln|\alpha|/\gamma$ the $\beta$-function can be simplified to:
\begin{equation}
	\label{eq:exponential_class_simplified}	
	\beta(\phi)=-\sqrt{6\lambda}-e^{-\gamma\phi}\;,
\end{equation}
and we can easily show that $\phi_\textrm{f} = - \ln\left[1 -\sqrt{6\lambda}\right]/\gamma$.\\

\begin{figure}[htb]
\centering
\includegraphics[width=1.\textwidth]{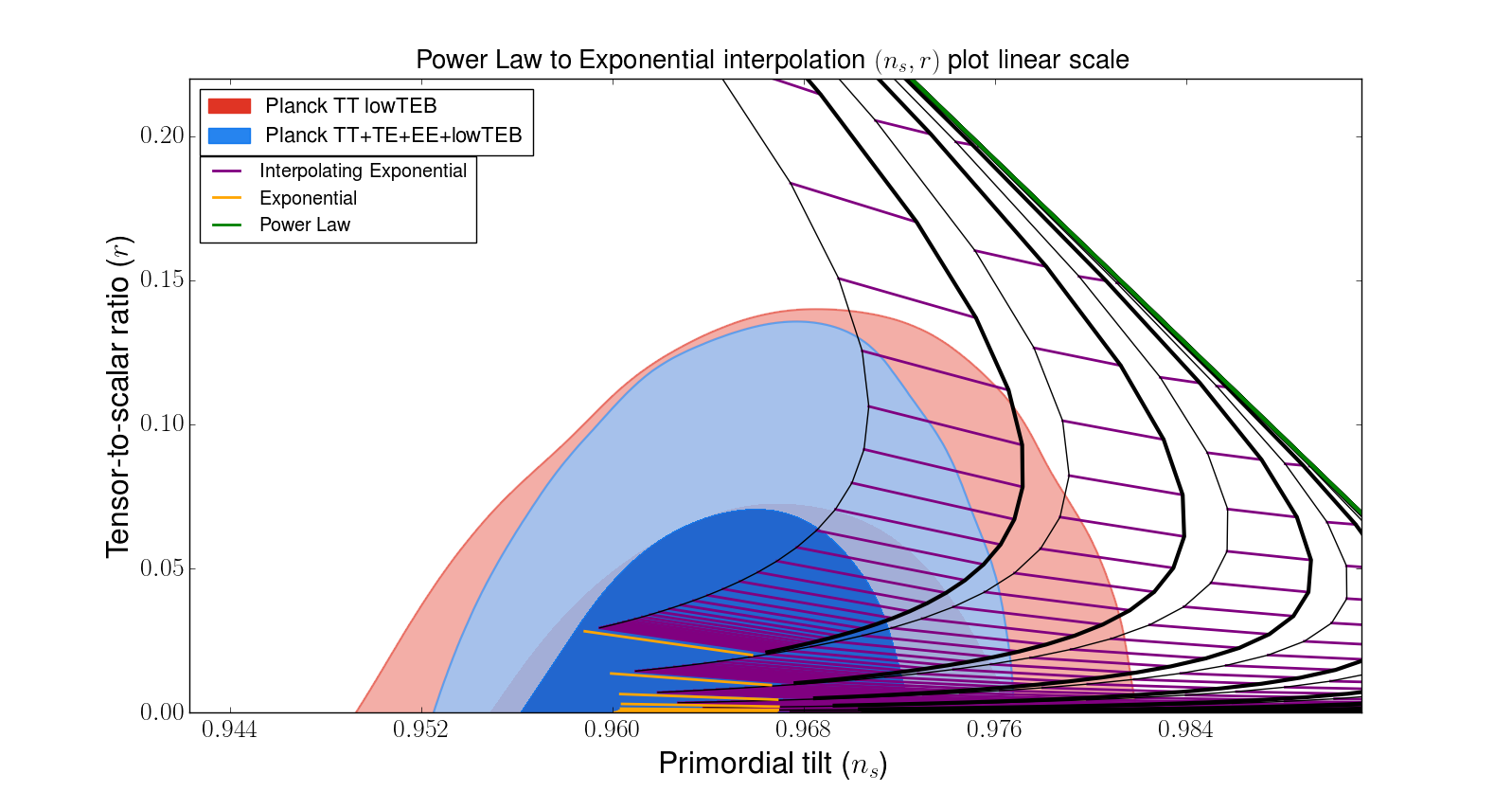}
\caption{Predictions for models described by the $\beta$-function of Eq.~\eqref{eq:exponential_class} (in purple) compared with Planck 2015 constraints~\cite{Ade:2015lrj} in the $(n_s,r)$ plane in linear scale. For comparison we show the predictions for the Exponential class (in orange) and for the Power law class (in green). In this plot we choose values for $\lambda$ in the interval $\lambda \in [10^{-6},0.8] $ and values of $\gamma$ in the interval $\gamma \in [0.3,10]$. Both the values of $\lambda$ and $\gamma$ are chosen with a even logarithmic spacing. Each horizontal segment corresponds to values of $N$ ranging from $50$ to $60$ (from left to right). More details on this plot are given in the main text.\label{fig:exponential}}
\end{figure}

\begin{figure}[htb]
\centering
\includegraphics[width=1.\textwidth]{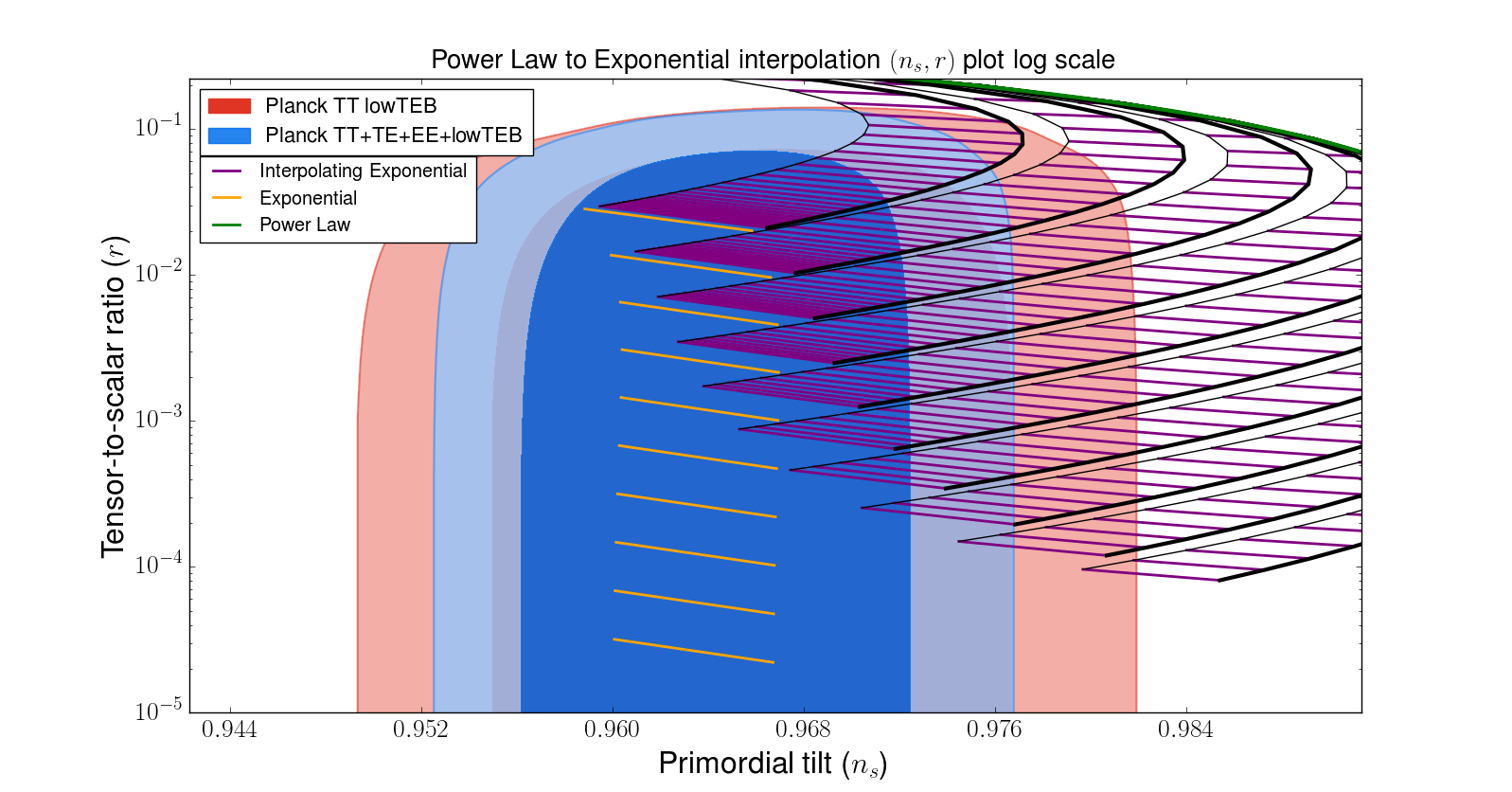}
\caption{Predictions for models described by the $\beta$-function of Eq.~\eqref{eq:exponential_class} (in purple) in the $(n_s,r)$ plane in semi-logarithmic scale. The values of $\lambda$ and $\gamma$ used for this plot are the same as in Fig.~\ref{fig:exponential}. \label{fig:exponentiallog}}

\end{figure}

This class of models corresponds to potentials that at the lowest order can be expressed as:
\begin{align}
	\label{eq:pot_exp_app}
	V(\phi)& \simeq V_\textrm{f} \left(1 - C e^{-\gamma\phi} \right) \exp\left\{\sqrt{6\lambda}\phi\right\}\;,
\end{align}
where $V_\textrm{f}$ is the normalization of the inflationary potential (that as usual can be set using the COBE normalization~\cite{Ade:2015xua,Ade:2015lrj}) and $C$ is a constant which can be expressed in terms of $\gamma$ and $\lambda$. Let us proceed by computing the expression for the number of e-foldings:
\begin{equation}
	\label{eq:efold_exp}
	N(\phi) =\frac{1}{\gamma\sqrt{6\lambda}}\ln\left(\frac{1+\sqrt{6\lambda}e^{\gamma\phi}}{1+\sqrt{6\lambda}e^{\gamma\phi_{\mathrm{f}}}}\right) =\frac{1}{\gamma\sqrt{6\lambda}}\ln\left[\left(1-\sqrt{6\lambda}\right)\left(1+\sqrt{6\lambda}e^{\gamma\phi}\right)\right] \; ,
\end{equation}
which is positive since $\phi$ decreases during inflation. We can now invert Eq.~\eqref{eq:efold_exp} in order to get $\phi(N)$:
\begin{equation}
	\label{eq:phitoN_exp}
	\phi(N)=\frac{1}{\gamma}\ln\left(\frac{e^{\gamma\sqrt{6\lambda}N}+\sqrt{6\lambda}-1}{\sqrt{6\lambda}-6\lambda}\right)\;,
\end{equation}
and similarly we can express the $\beta$-function in terms of $N$:
\begin{equation}
	\label{eq:beta_N}
	\beta(N)=-\frac{\sqrt{6\lambda}e^{\gamma\sqrt{6\lambda}N}}{e^{\gamma\sqrt{6\lambda}N}+\sqrt{6\lambda}-1} \; .
\end{equation}
It is important to stress that despite the rather complicated form of the potential shown in Eq.~\eqref{eq:pot_exp_app}, the corresponding $\beta$-function is simple and convenient to manipulate. In particular we have shown that the model can be completely solved analytically. This is a clear difference with respect to the usual approach where it is quite difficult to specify potentials that lead to an evolution that admits analytical solutions. This illustrates the strength of this approach which allows  constructing new interesting cases that can be solved analytically (and that eventually lead to interesting predictions for cosmological observables) even if they correspond to non-trivial potential. Finally, the predictions for $n_s$ and $r$ in terms of $N$ can be directly obtained from Eq.~\eqref{eq:general_nsandr}. The results for a set of models in this class are shown in the plots on Figs.~\ref{fig:exponential} and Fig.~\ref{fig:exponentiallog} (in linear and semi-logarithmic scale respectively). Before discussing these plots in detail let us describe the two asymptotic limits $\gamma\sqrt{6\lambda}N\ll 1$ and $\gamma\sqrt{6\lambda}N\gg 1$ which are respectively reproducing the Exponential and Power Law classes. \\


In the limit $\gamma\sqrt{6\lambda}N\ll 1$ we can easily get: 
\begin{equation}
\phi(N)\simeq \frac{1}{\gamma}\ln\left(\frac{1+\gamma N}{1-\sqrt{6\lambda}}\right) \; , \qquad \text{and} \qquad \beta(N)\simeq -\frac{1}{1+\gamma N} \; ,
\end{equation}
which is precisely the $\beta$-function of the Exponential class. Conversely, in the limit $\gamma\sqrt{6\lambda}N\gg 1$ we find:
\begin{equation}
\phi(N)\simeq \sqrt{6\lambda}N-\frac{1}{\gamma}\ln(\sqrt{6\lambda}-6\lambda) \; ,  \qquad \text{and} \qquad \beta(N) \simeq -\sqrt{6\lambda} \; ,
\end{equation}
being $e^{-\gamma\sqrt{6\lambda}N}\simeq 0$. We see that in this limit we recover the Power Law class.\\

In the plots of Fig.~\ref{fig:exponential} and Fig.~\ref{fig:exponentiallog} we show the predictions for $n_s$ and $r$ (respectively in linear and semi-logarithmic scale) for the models described in this section. Notice that in this case the $\beta$-function is completely specified by the value of the two critical exponents $\gamma$ and $\lambda$. Once these two parameters are fixed, we can compute numerical predictions as a function of $N$ only. In order to produce the plots of Fig.~\ref{fig:exponential} and Fig.~\ref{fig:exponentiallog}, we first fix some values of $\gamma$ (with even log spacing from $\gamma =0.3$ to $\gamma = 10$) and varying the value of $\lambda$ (again with even log spacing from $\lambda = 0.8$ to $\lambda = 10^{-6}$) we show the interpolation between the two asymptotic limits \emph{i.e.} the Exponential (in orange) and the Power law (in green). The solid black lines are used to follow the variation of $\lambda$ while keeping the values of $\gamma$ and $N$ fixed. The thick black line corresponds to $ N = 60 $ and the thin black line corresponds to $N = 50$. As for larger values of $\gamma$ the exponential term approaches zero more rapidly (going deeper into the inflationary phase), smaller values of $\lambda$ are required in order to approach the exponential limit at $N\simeq 50 \div 60$. This feature is manifest in the plot of Fig.~\ref{fig:exponentiallog}. Notice that between the two asymptotic limits there is an intermediate region with a whole new set of inflationary models.

\subsubsection{Inverse}
\label{sec:inverse}
In this section we consider the inverse of a monomial as a correction to the constant $\beta$-function\footnote{Note that as already stated in Sec.~\ref{sec:positive}, in order to be consistent with the convention used in~\cite{Binetruy:2014zya,Pieroni:2015cma}, the $\beta$-function is taken to be negative and the field $\phi$ is positive. An analogous derivation can be carried out with a positive $\beta$-function and a negative valued field by adjusting the sign of the parameter $\alpha$.} ($\alpha >0$):
\begin{equation}
	\label{eq:inverse_class}
	\beta(\phi) = -\sqrt{6 \lambda} - \frac{\alpha}{\phi^n}  \;.
\end{equation}
This case is expected to interpolate between a power law and either the chaotic class (for $n=1$) or the inverse class ($n >1$). This behavior will be shown explicitly for the cases $n =1$ and $n = 2 $ where analytical expressions for $N(\phi)$ can be obtained. Before considering the two cases separately, it is useful to compute the general expression for $\phi_{\textrm{f}}$ using $|\beta(\phi_{\textrm{f}})| =1 $:
\begin{equation}
	\phi_{\textrm{f}} = \left( \frac{\alpha}{ 1 -\sqrt{6 \lambda}} \right)^{1/n} \; .
\end{equation}
Notice that analogously to the Exponential case discussed in the previous section, this is a large field model where inflation occurs for $\phi_{\textrm{f}} \lesssim \phi $. At lowest order the potential associated with this model is:
\begin{align}
	V(\phi)& \simeq V_\textrm{f} \left( 1 - \frac{\alpha}{n-1} \phi^{1-n} \right) e^{\sqrt{6\lambda}\phi} \;,&&n > 1\;,\\
	V(\phi)&\simeq V_\textrm{f} \, \phi^\alpha e^{\sqrt{6\lambda}\phi}\;,&&n=1\;.
\end{align}
Let us now focus on the two cases $n=1$ and $n=2$.

\begin{itemize}
	\item \textbf{Chaotic,  $n = 1$}\\
	In this case we easily get:
\begin{equation}
	\label{eq:inverse_class_N_n1}
	N(\phi) = \frac{\phi - \phi_\textrm{f}}{\sqrt{6 \lambda}} - \frac{\alpha}{6 \lambda } \ln \left[ \frac{1 + \frac{\sqrt{6 \lambda}}{\alpha} \phi}{1 + \frac{\sqrt{6 \lambda}}{\alpha} \phi_{\textrm{f}}} \right] \; .
\end{equation}
In general this equation cannot be inverted to get $\phi(N)$. However, we can show that there are two asymptotic behaviors. As a first step we multiply both sides by $6\lambda / \alpha$ and define $\sqrt{6 \lambda} \phi  / \alpha \equiv z$, we then express Eq.~\eqref{eq:inverse_class_N_n1} as:
\begin{equation}
	\label{eq:equation_for_N_inverse_n1}
	\frac{6\lambda N}{\alpha} = z - z_{\textrm{f}} -  \ln \left[ \frac{1 + z }{1 + z_{\textrm{f}}} \right] \; .
\end{equation}
Notice that as $z > z_{\textrm{f}}$, the second term on the r.h.s. of this equation is negative. At this point we can consider the two limiting cases:
\begin{enumerate}
	\item In the limit $ \frac{6\lambda N}{\alpha} \ll 1$ a solution for Eq.~\eqref{eq:equation_for_N_inverse_n1} exists for $z \ll 1$. We approximate the logarithm on the right-hand side to get:
	\begin{equation}
		\frac{6\lambda N}{\alpha}  \simeq z^2 /2 - z_{\textrm{f}}^2 /2 \;.
	\end{equation}
	We can thus express the field as function of $N$ as:
	 \begin{equation}
	 	\phi(N) \simeq \sqrt{2 \alpha N + \phi_{\textrm{f}}^2}\;,
	 \end{equation}
	where $\phi_{\textrm{f}}=\alpha(1-\sqrt{6\lambda})^{-1}$. The $\beta$-function becomes:
	 \begin{equation}
	 	\beta(N) \simeq -\sqrt{6\lambda} - \alpha /\sqrt{2 \alpha N + \phi_{\textrm{f}}^2} \simeq -   \sqrt{ \frac{\alpha}{2 N} }\;,
	 \end{equation}
	 and we recognize the expression for $\phi(N)$ of the Chaotic class \textbf{Ib(1)} (see Sec.~\ref{subsubsec:General_Exact_solution_exact_chaotic}). For completeness we report the predictions for $n_s$ and $r$:
	 \begin{equation}
	 	n_s - 1 \simeq - \frac{1 + \alpha/2}{N}  \;, \qquad r \simeq \frac{4 \alpha}{N} \; .
	 \end{equation}
	\item In the limit $\frac{6\lambda N}{\alpha} \gg 1$ the solution for Eq.~\eqref{eq:equation_for_N_inverse_n1} exists for $z \gg 1$. The linear term dominates on the right-hand side and we get:
	\begin{equation}
		\phi(N) = \sqrt{6 \lambda} N + \phi_\textrm{f} \; .
	\end{equation}
	At lowest order the $\beta$-function as a function of $N$ is simply a constant and thus corresponds to the power law class \textbf{Ib(0)} (see Sec.~\ref{subsubsec:General_Exact_solution_cst}).
\end{enumerate}

\begin{figure}[htb]
\centering
\includegraphics[width=1.\textwidth]{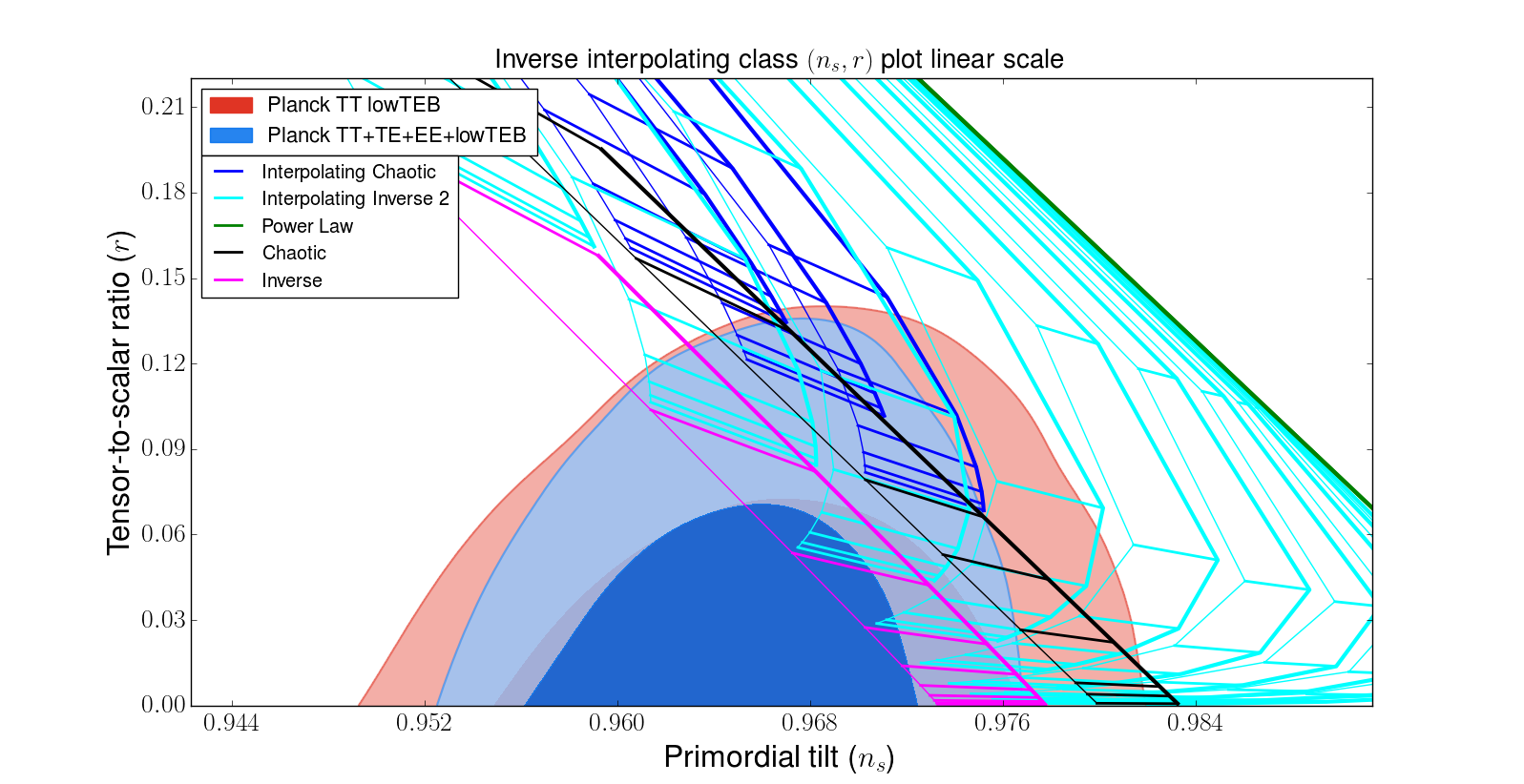}
\caption{Prediction for models described by the $\beta$-function of Eq.~\eqref{eq:inverse_class} compared with Planck 2015 constraints~\cite{Ade:2015lrj} in the $(n_s,r)$ plane in linear scale. We show the results both for $n=1$ (in blue) and for $n=2$ (in teal). For comparison we are also showing the predictions of the power law class (in green), the chaotic class (in black) and the inverse class with $p=2$ (in magenta). For $n=1$ we use $\alpha = 0.33,0.5,1,2,3$ and $\lambda \in [10^{-6},10^{-1}]$ with even logarithmic spacing, for $n=2$ we use values for $\alpha \in [10^{-2},10^2]$ and $\lambda \in [10^{-6},10^{-1}]$ both  with even logarithmic spacing. Horizontal segment corresponds to $N \in [50,60]$ (from left to right).}
\label{fig:inverse}
\end{figure}

\item \textbf{Quadratic inverse, n = 2}\\
	In this case the number of e-foldings reads:
\begin{equation}
	\label{eq:inverse_class_N_n2}
	N(\phi) = \frac{\phi - \phi_\textrm{f}}{\sqrt{6 \lambda}} - \sqrt{\frac{\alpha}{6 \lambda\sqrt{6 \lambda} }}\left[  \arctan \left(\sqrt{\frac{\sqrt{6 \lambda}}{\alpha}} \phi \right)   -  \arctan \left(\sqrt{\frac{\sqrt{6 \lambda}}{\alpha}}\phi_{\textrm{f}} \right) \right] \; .
\end{equation}
In general this equation cannot be inverted to get $\phi(N)$ but again we can consider the two asymptotic behaviors. We start by multiplying both sides by $(6\lambda)^{3/4} / \sqrt{\alpha}$ and define $(6 \lambda)^{1/4}\phi / \sqrt{\alpha}  \equiv z$ so that Eq.~\eqref{eq:inverse_class_N_n2} reads:
\begin{equation}
	\label{eq:equation_for_N_inverse_n2}
	\frac{(6 \lambda)^{3/4} N}{\sqrt{\alpha}} = z - z_{\textrm{f}} -   \left[\arctan \left(z \right) - \arctan \left( z_{\textrm{f}}\right) \right] \; .
\end{equation}
We then notice that as $z > z_{\textrm{f}}$ the second term on the r.h.s. is always negative and we proceed by distinguishing the two limiting cases:
\begin{enumerate}
	\item In the limit $ \frac{(6 \lambda)^{3/4} N}{\sqrt{\alpha}} \ll 1$ there is a solution for Eq.~\eqref{eq:equation_for_N_inverse_n2} when $z \ll 1$. In this limit we can Taylor expand the $\arctan$ to get:
	\begin{equation}
		\frac{(6\lambda)^{3/4} N}{\sqrt{\alpha} }  \simeq \frac{z^3}{3} - \frac{z_{\textrm{f}}^3}{3}\;,
	\end{equation}
	 and therefore:
	 \begin{equation}
	 	\phi(N) \simeq\left( 3 \alpha N + \phi_\textrm{f}^3 \right)^{\frac{1}{3}}\;.
	 \end{equation}
	 Substituting into Eq.~\eqref{eq:inverse_class} we get:
	 \begin{equation}
	 	\beta(N) = -\sqrt{6 \lambda} - \alpha \left( 3 \alpha N + \phi_\textrm{f}^3 \right)^{-\frac{2}{3}} \simeq - \left( \frac{\sqrt{\alpha}}{3 N}\right)^{2/3} \;,
	 \end{equation}
	that corresponds to the Inverse class \textbf{Ib(p)} with $n=2$. For completeness the predictions for $n_s$ and $r$ are:
	\begin{equation}
		n_s - 1 \simeq -\frac{4}{3 N} \; , \qquad r \simeq \frac{8 \alpha^{2}{3}}{ (3 N)^{4/3}} \;.
	\end{equation}

	\item In the limit $\frac{(6 \lambda)^{3/4} N}{\sqrt{\alpha}} \gg 1$, we must look for a solution of Eq.~\eqref{eq:equation_for_N_inverse_n2} with $z \gg 1$. In this case the linear term dominates the r.h.s and we simply get:
	\begin{equation}
		\phi(N) = \sqrt{6\lambda}N+ \phi_\textrm{f}  \; .
	\end{equation}
	That again corresponds to the power law class \textbf{Ib(0)} (see Sec.~\ref{subsubsec:General_Exact_solution_cst}).
\end{enumerate}
\end{itemize}

In the plot of Fig.~\ref{fig:inverse} we show the predictions for $n_s$ and $r$ (in linear scale) for the class of models described in this section. The procedure used to produce this plot is similar to the one defined in Sec.~\ref{sec:exponential}. We can see that (as expected) for small values of $\lambda$ the predictions for the classes of models are approaching the predictions for the chaotic and the inverse class respectively. On the other hand for larger values of $\lambda$ we recover the usual power law behavior. Notice that in this plot (and more generally in this section) we are only considering models for which we can perform analytical computations. However, according to the discussion of~\cite{Binetruy:2014zya}, for larger values of $n$ we can define models of the inverse class which lead to predictions for $n_s$ and $r$ that are in even better agreement with the experimental constraints. Clearly, these models could be used to implement the mechanism discussed in this paper leading to the definition of models which predict better values for $n_s$ and $r$.

\subsubsection{Monomial}
\label{sec:monomial}
In this section we consider a $\beta$-function of the form:
\begin{equation}
	\label{eq:monomial_class}
	\beta(\phi) = \sqrt{6\lambda} + \alpha \phi^n\;,
\end{equation}
with $\alpha>0$. Conversely to the cases discussed in Sec.~\ref{sec:exponential} and in Sec.~\ref{sec:inverse}, for this class of models inflation is realized for $\phi \rightarrow 0$. During this epoch the field $\phi$ is positive and is monotonically increasing from $\phi=0$ to $\phi_f = [ ( 1 - \sqrt{6 \lambda})/\alpha ]^{1/n}$. Notice that for $\alpha > 1$ the field excursion during inflation is smaller than $1$. \\

As a first step we show that the case $n=1$ can be safely ignored. Since in this case the $\beta$-function is simply $\beta(\phi) = \sqrt{6\lambda} + \alpha \phi$, a field redefinition $\phi = \phi' - \sqrt{6\lambda}/\alpha $ leads simply to the usual monomial class with $\beta(\phi') = \alpha \phi'$. Analogously, for any odd power of $n$ we can perform a field redefinition $\phi = \phi' - (\sqrt{6\lambda}/\alpha)^{1/n}$ to get:
\begin{equation}
	\beta(\phi) = \sum_{i = 1}^n c_i \phi^i\;,
\end{equation}
where the $c_i$ denote some constant factors. As the constant term disappears, these cases are already discussed in~\cite{Binetruy:2014zya} and they are not relevant for the analysis presented in this paper. For this reason we can simply restrict to cases with an even value for $n$. \\

The potentials for these models (with $n>1$) are given by: 
\begin{align}
	V(\phi)&=V_\textrm{f} \, \exp\left\{-\sqrt{6\lambda}\phi-\frac{\alpha}{n+1}\phi^{n+1}\right\}\left[1-\lambda-\frac{1}{6}\left(2\sqrt{6\lambda}\alpha\phi^n+\alpha^2\phi^{2n}\right)\right] \, .
\end{align} 
As we explain in the following, in order to provide a consistent model of inflation the field excursion is non-trivially related to the values of $\alpha$ and $\lambda$. Therefore we cannot provide a general first order expression for the potential. \\ 


\begin{figure}[htb]
\begin{center}
\centering
\includegraphics[width=1.\textwidth]{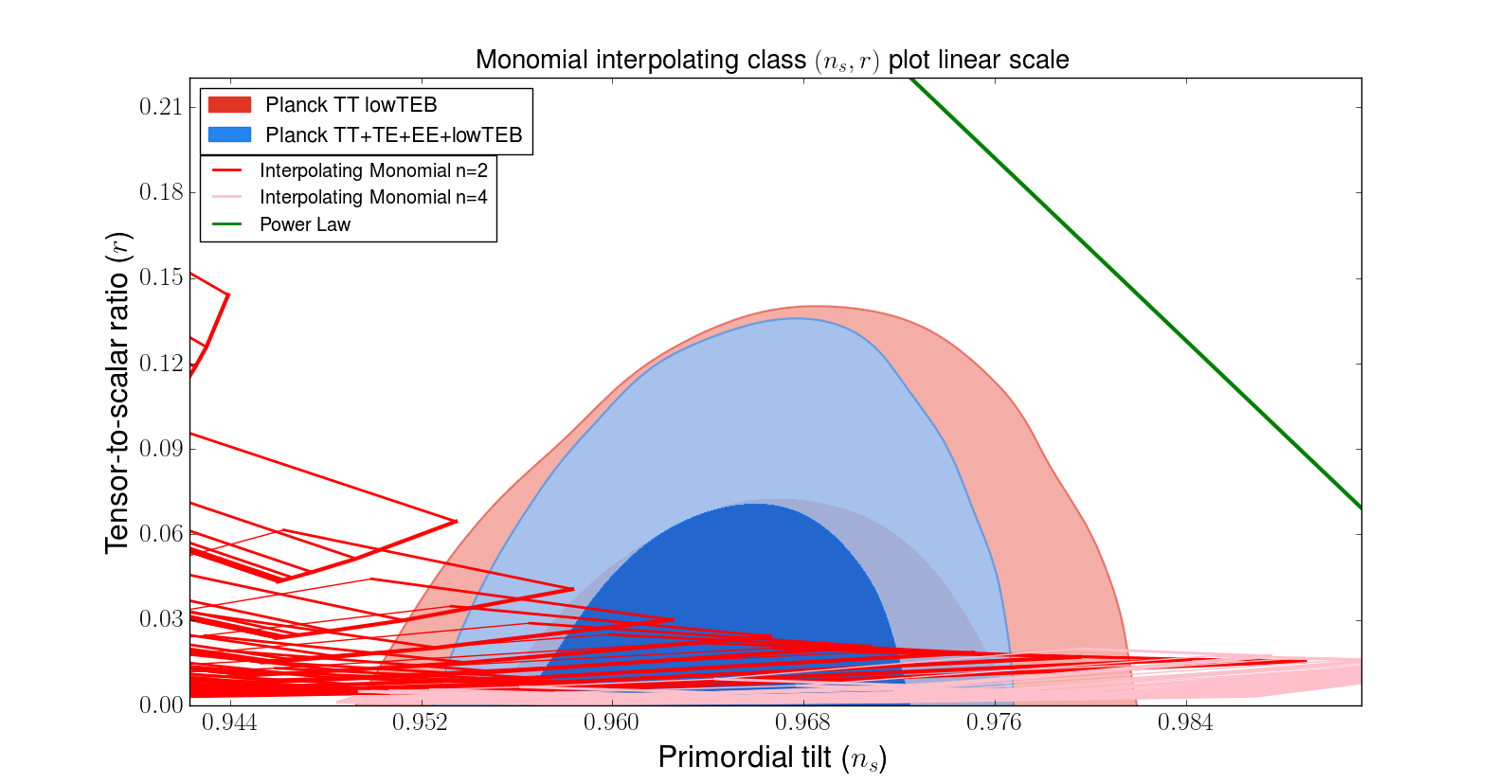}
\end{center}
\caption{Prediction for models described by the $\beta$-function of Eq.~\eqref{eq:monomial_class} compared with Planck 2015 constraints~\cite{Ade:2015lrj} in the $(n_s,r)$ plane in linear scale. We show the results both for $n=2$ (in red) and for $n=4$ (in pink) as well as for the power law class (in green). For $n=2$ we used values for $\lambda$ in the interval $\lambda \in [10^{-10},10^{-3.5}] $ and values of $\alpha$ in the interval $\alpha \in [10^{-3},10^{-2}]$. For $n=4$ we used values for $\lambda$ in the interval $\lambda \in [10^{-10},10^{-3.5}] $ and values of $\alpha$ in the interval $\alpha \in [10^{-4},10^{-3}]$. Both the values of $\lambda$ and $\alpha$ are chosen with a even logarithmic spacing. Horizontal lines corresponds to values of $N$ ranging from $50$ to $60$ (from left to right).}
\label{fig:monomial}
\end{figure}


In order to discuss an explicit example which can be solved fully analytically let us focus on the case with $n=2$. The first step is to compute the number of e-foldings:
\begin{equation}
	N =-\frac{1}{\sqrt{\alpha\sqrt{6\lambda}}}\left(\arctan\left[\sqrt{\frac{\alpha}{\sqrt{6\lambda}}}\phi\right]-\arctan\left[\sqrt{\frac{1-\sqrt{6\lambda}}{\sqrt{6\lambda}}}\right]\right)\;,
\end{equation}
where we have substituted $\phi_{\textrm{f}}^2=(1-\sqrt{6\lambda})/\alpha$. Before proceeding further, it is interesting to stress that these models can support inflation only for a limited number of e-foldings. This can be easily checked by computing the value of $N$ for $\phi = 0$:
\begin{align}
	N_{\text{tot}}&=N(\phi=0)=\frac{1}{\sqrt{\alpha\sqrt{6\lambda}}}\arctan\left[\sqrt{\frac{1-\sqrt{6\lambda}}{\sqrt{6\lambda}}}\right]\;.
\end{align}
For the model to be a realistic description of inflation, we then need $N_{\text{tot}}$ to be larger than $50 \div 60$. This will turn into a constraint for the maximal value $\alpha$ at a given value of $\lambda$. For example if we require $N_{\text{tot}} \gtrsim 60$ we get:
\begin{equation}
	\label{eq:monomial_alpha_max}
	\alpha \lesssim \frac{1}{3600\sqrt{6\lambda}}\arctan^2\left[\sqrt{\frac{1-\sqrt{6\lambda}}{\sqrt{6\lambda}}}\right]\;.
\end{equation}
We can proceed by computing the expression for $\phi$ as a function of $N$:
\begin{align}
	\label{eq:monomial_efold}
	\phi(N)&=\sqrt{\frac{\sqrt{6\lambda}}{\alpha}}\tan\left[-\sqrt{\alpha\sqrt{6\lambda}}N+\arctan\left[\sqrt{\frac{1-\sqrt{6\lambda}}{\sqrt{6\lambda}}}\right]\right]\;.
\end{align}
 Notice that the r.h.s. of this equation is positive \emph{i.e.} the ``fixed point'' $\phi = 0$ is consistently approached for $N = N_{\text{tot}}$. Moreover, it is interesting to notice that if we choose the parameter $\alpha$ to be at its maximal value given by~\eqref{eq:monomial_alpha_max}, we have $\phi = 0$ for $N_{\text{tot}}=60$. In this case the model would have an overall inflationary period of exactly $60$ e-foldings and thus the predictions at $N=60$ are given by the constant $\sqrt{6\lambda}$ \emph{i.e.} by the power law fixed point\footnote{Notice however that in order to impose the Bunch-Davies vacuum for the perturbations that leave the horizon at $N = 60$ we need $N_{\text{tot}} > 60$.}. \\

Using Eq.~\eqref{eq:monomial_efold} the $\beta$-function can be expressed as:
\begin{equation}
	\beta(N) = \sqrt{6\lambda}\left(1+\tan^2\left[-\sqrt{\alpha\sqrt{6\lambda}}N+\arctan\left[\sqrt{\frac{1-\sqrt{6\lambda}}{\sqrt{6\lambda}}}\right]\right]\right)\;,
\end{equation}
again $n_s$ and $r$ are respectively given by:
\begin{align}
	n_s-1&=-\left[\beta^2(N)+2\beta_{,\phi}(N)\right] \; , \qquad r=8\beta^2(N)\;,
\end{align}
where $\beta_{,\phi}(N)$ is obtained by taking first the $\phi$ derivative of Eq.~\eqref{eq:monomial_class} and then substituting Eq.~\eqref{eq:monomial_efold}.\\

The plot of Fig.~\ref{fig:monomial} shows the predictions for $n_s$ and $r$ (in linear scale) for the class of models described in this section\footnote{Again the procedure used to obtain these plots is similar to the one defined in Sec.~\ref{sec:exponential}}. Notice that in this plot we are also showing the predictions for a model $(n=4)$ where we are not able to perform a fully analytical treatment. Interestingly both the case with $n=2$ and $n=4$ predict values for $n_s$ and $r$ that are in agreement with the constraints set by the latest cosmological data~\cite{Ade:2015xua,Ade:2015lrj,Array:2015xqh}. Even in this case we can appreciate the interpolating behavior of the $\beta$-function. If $\sqrt{6\lambda}\ll1$ we simply recover the monomial case, and no difference can be appreciated in the last $60$ e-foldings. Conversely, for larger values of $\lambda$ (but still in agreement with the constraint of at least $50 \div 60$ e-foldings), we approach power law predictions. In summary, we have defined a set of inflationary models that are in agreement with the latest cosmological data and that lead to a finite period of inflation.

\section{Conclusions and outlook}
\label{sec:conclusions}
In this work we have studied constant-roll inflation in terms of the $\beta$-function formalism. In particular we have shown that the constant-roll condition translates into a simple first order differential equation for $\beta(\phi)$. We have derived the solutions of this equation and shown that they reproduce the constant-roll models already studied in the literature. Interestingly, among the cases discussed in this paper, we have recognized some $\beta$-functions that were already considered in the original work on this topic~\cite{Binetruy:2014zya}. This is a consequence of the generality of the results obtained in terms of this formalism (which allows performing analytical computations even beyond the usual slow-roll approximation). Finally we have discussed the interpolating behavior of some solutions. This first part of the work is intended to show the simplicity of the $\beta$-function formalism when dealing with concrete examples. \\

Having reproduced the models that are already known in the literature, we have also shown that this formalism is convenient to go beyond exact solutions. In particular we have defined a set of approximate solutions for the constant-roll equation that correspond to models with phenomenological predictions in good agreement with the latest experimental data. With this procedure we have been able to consider some new classes of models that to our knowledge are not present in the literature. For completeness (even if our approach is not based on the potential) we derived the form of the scalar potential associated with each model. The rather complicated form of the potentials in balance with the relative ease to perform analytical computations illustrates once again the strength of the method. \\

We have considered in greater details the power-law solution and its different extensions. It is well known that an exact power-law model of inflation is incomplete since there is no mechanism to end the period of inflation. We showed that by adding a correction to the corresponding $\beta$-function, it becomes possible to have concrete models with a natural end to the inflationary period. For the specific case of a monomial correction to the $\beta$-function, we found the interesting feature that the total period of inflation may only be finite. Most of the asymptotic power-law models studied in this work predict cosmological parameters that are in agreement with the most recent observational constraints~\cite{Ade:2015xua,Ade:2015lrj,Array:2015xqh} which is an interesting result from a phenomenological point of view. An intriguing subject for further works on this topic would be the search for peculiar observable signatures that allows distinguishing these models from standard dS asymptoting solutions. \\

In this work we illustrated the strength of the $\beta$-function approach when dealing with concrete and phenomenologically interesting models. An interesting prospect for future works would be to relate these models with some theoretically well-motivated theories (which could arise for example from the usual QFT approach). As already stated in Sec.~\ref{sec:positive}, the holographic interpretation might play a significant role in this analysis. The three dimensional QFTs that are dual to asymptotic power-law models are theories with generalized conformal invariance~\cite{Jevicki:1998ub} which are a particular case (with $z = 1$ ) of theories that violate hyperscaling\footnote{In these theories the so-called hyperscaling violation exponent $\theta$ can be introduced in order to quantify the degree of violation of scale invariance.}~\cite{Jevicki:1998ub,Charmousis:2010zz,Gouteraux:2011ce,Huijse:2011ef,Kanitscheider:2008kd,Dong:2012se,Alishahiha:2012cm,Gath:2012pg,Hartnoll:2016apf}. These models are thus conceptually different from dS asymptoting solutions whose holographic dual theories are deformed CFTs. As a consequence, while phenomenologically power-law models are quite similar to (and in some cases nearly indistinguishable from) the most common realization of inflation, there are deep theoretical differences between these two classes of models.\\

Asymptotic power-law solutions are not only interesting in the context of inflationary model building but they can also be useful while studying the late time evolution of our Universe. In particular a wide set of models with exponentially flat potentials (which lead to a power-law expansion) has been studied in the literature~\cite{Capozziello:2005ra,Rubano:2001su,Rubano:2003et,Pavlov:2001dt,Rubano:2002mc,Demianski:2004qt}. The $\beta$-function formalism is not only suitable to describe inflation but rather it can also be used to describe the late time evolution of Universe driven by quintessence\footnote{A classification of models of quintessence in term of the $\beta$-function formalism was carried out in~\cite{Cicciarella:2016dnv}.}. In this case the situation is typically reversed and the period of accelerated expansion corresponds to an RG flow towards a fixed point. As discussed in~\cite{Cicciarella:2016dnv}, the presence of matter naturally induces a flow and thus in this case a consistent late time evolution can be attained even if the $\beta$-function is exactly constant.\\

While in this work we have only focused on the case of a single scalar field with standard derivative term and minimal coupling to gravity, it has been already shown in~\cite{Pieroni:2015cma,Binetruy:2016hna} that the $\beta$-function formalism can be consistently extended to more general models of inflation. However, it is fair to point out that generalizations of the formalism in order to include for example multi-fields models of inflation, non-minimal derivative couplings between the inflaton and gravity\footnote{As in the case of ``new Higgs'' inflation~\cite{Germani:2010gm,Germani:2010ux}.} and models of modified gravity are still lacking. The analysis carried out in~\cite{Bourdier:2013axa} and~\cite{Garriga:2014fda,Garriga:2015tea} can provide a useful guideline in order to define the first of these generalizations.

\vspace{1cm}
\subsubsection*{Acknowledgements}

We thank V.~Domcke, E.~Kiritsis and D.~Langlois for their valuable comments and for very useful and helpful discussions. M.P. would like to thank the Niels Bohr Institute and Paris Center for Cosmological Physics for hospitality at various stages of this work.
J.M. is supported by Principals Career Development Scholarship and Edinburgh Global Research Scholarship. M.P. acknowledges the support of the Spanish MINECO's ``Centro de Excelencia Severo Ochoa'' Programme under grant SEV-2012-0249. This project has received funding from the European Union’s Horizon 2020 research and innovation programme under the Marie Sk\l{}odowska-Curie grant agreement No 713366.

\vspace{1cm}

\providecommand{\href}[2]{#2}\begingroup\raggedright

\end{document}